\documentclass[journal=jpclcd,manuscript=article]{achemso}
\usepackage[utf8]{inputenc}
\usepackage{amsmath}
\usepackage{physics}
\usepackage{graphics}
\usepackage{url}
\usepackage{textcomp}
\usepackage[hidelinks]{hyperref}
\usepackage{soul}


\newcommand{\micron}{$\,\rm \mu m$}
\newcommand{\mob}{$\,\rm cm^{2}\rm V^{-1}s^{-1}$}
\newcommand{\fluence}{\,$\mu\rm J\rm cm^{-2}$}
\newcommand{\mapbi}{MAPbI$_3$}
\newcommand{\mapbifull}{CH$_3$NH$_3$PbI$_3$}

\newcommand*{\citen}[1]{%
	\begingroup
	\romannumeral-`\x % remove space at the beginning of \setcitestyle
	\setcitestyle{numbers}%
	\cite{#1}%
	\endgroup   
}

\author{Chelsea Q. Xia}
\affiliation{Department of Physics, University of Oxford, Clarendon Laboratory, Parks Road, Oxford OX1 3PU, U.K.}

\author{Jiali Peng}
\affiliation{Key Lab of Artificial Micro- and Nano-Structures of Ministry of Education of China, School of Physics and Technology, Wuhan University, Wuhan 430072, P. R. China}

\author{Samuel Ponc\'e}
\affiliation{Department of Materials, University of Oxford, Parks Road, Oxford OX1 3PH, U.K.}
\alsoaffiliation{Theory and Simulation of Materials (THEOS), \'Ecole Polytechnique F\'ed\'erale de Lausanne, CH-1015 Lausanne, Switzerland}

\author{Jay B. Patel}
\affiliation{Department of Physics, University of Oxford, Clarendon Laboratory, Parks Road, Oxford OX1 3PU, U.K.}

\author{Adam D. Wright}
\affiliation{Department of Physics, University of Oxford, Clarendon Laboratory, Parks Road, Oxford OX1 3PU, U.K.}

\author{Timothy W. Crothers}
\affiliation{Department of Physics, University of Oxford, Clarendon Laboratory, Parks Road, Oxford OX1 3PU, U.K.}

\author{Mathias Uller Rothmann}
\affiliation{Department of Physics, University of Oxford, Clarendon Laboratory, Parks Road, Oxford OX1 3PU, U.K.}

\author{Juliane Borchert}
\affiliation{Department of Physics, University of Oxford, Clarendon Laboratory, Parks Road, Oxford OX1 3PU, U.K.}

\author{Rebecca L. Milot}
\affiliation{Department of Physics, University of Warwick, Gibbet Hill Road, Coventry CV4 7AL, U.K.}

\author{Hans Kraus}
\affiliation{Department of Physics, University of Oxford, Denys Wilkinson Building, Keble Road, Oxford OX1 3RH, U.K.}

\author{Qianqian Lin}
\affiliation{Key Lab of Artificial Micro- and Nano-Structures of Ministry of Education of China, School of Physics and Technology, Wuhan University, Wuhan 430072, P. R. China}

\author{Feliciano Giustino}
\affiliation{Department of Materials, University of Oxford, Parks Road, Oxford OX1 3PH, U.K.}
\alsoaffiliation{Oden Institute for Computational Engineering and Sciences, University of Texas at Austin, Austin, Texas 78712, USA}
\alsoaffiliation{Department of Physics, University of Texas at Austin, Austin, Texas 78712, USA}

\author{Laura M. Herz}
\affiliation{Department of Physics, University of Oxford, Clarendon Laboratory, Parks Road, Oxford OX1 3PU, U.K.}

\author{Michael B. Johnston}
\email{michael.johnston@physics.ox.ac.uk}
\affiliation{Department of Physics, University of Oxford, Clarendon Laboratory, Parks Road, Oxford OX1 3PU, U.K.}

\title[An \textsf{achemso} demo]
{Limits to Electrical Mobility in Lead-Halide Perovskite Semiconductors}

\begin{document}
	
	\begin{tocentry}
		\includegraphics{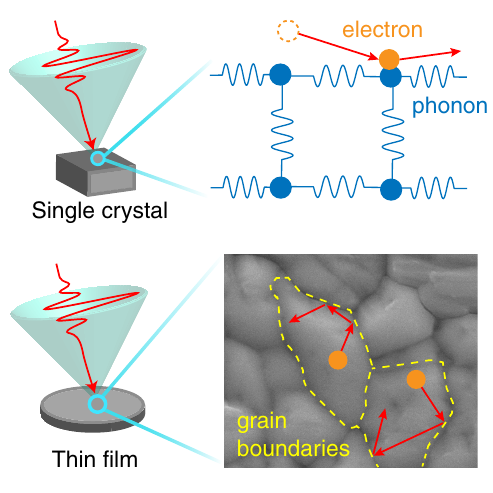}
	\end{tocentry}

\begin{abstract}
	Semiconducting polycrystalline thin films are cheap to produce and can be deposited on flexible substrates, yet high-performance electronic devices usually utilize single-crystal semiconductors, owing to their superior electrical mobilities and longer diffusion lengths. Here we show that the electrical performance of polycrystalline films of metal-halide perovskites (MHPs) approaches that of single crystals at room temperature. Combining temperature-dependent terahertz conductivity measurements and ab initio calculations we uncover a complete picture of the origins of charge scattering in single crystals and polycrystalline films of \mapbifull. We show that Fr\"ohlich scattering of charge carriers with multiple phonon modes is the dominant mechanism limiting mobility, with grain-boundary scattering further reducing mobility in polycrystalline films. We reconcile the large discrepancy in charge diffusion lengths between single crystals and films by considering photon reabsorption. Thus, polycrystalline films of MHPs offer great promise for devices beyond solar cells, including transistors and modulators.
\end{abstract}

In the last decade, organic-inorganic metal-halide perovskite (MHP) semiconductors have emerged as promising materials for photovoltaic applications,\cite{efficiency,application1,application2,application3} owing to their ease of large-scale deposition and excellent optoelectronic properties, such as high charge-carrier mobility,\cite{mobility_THz_christian1,mobility_THz_christian2,mobility_THz_becky} long carrier diffusion length\cite{diffusion1,hall_dong2015electron,sclc_shi2015low} and composition-tunable band gap.\cite{bandgap1,bandgap2} To date the primary application of these materials has been photovoltaics. In particular, single-junction solar cells based on these materials have shown rapid growth in solar to electrical power conversion efficiency (PCE) from 3.8\% to 25.2\%, whereas perovskite-silicon tandem solar cells have reached efficiencies of over 29\%.\cite{efficiency,patel2020light} MHP single crystals have attracted intense interest due to their potentials in fabricating photodetectors,\cite{lian2015high} X-ray scintillators and detectors,\cite{wei2020all,shrestha2017high,peng2021crystallization} as well as applications in photocatalysis and photoelectrochemical fields.\cite{wang2021surface} On the other hand, the cutting-edge photovoltaic devices have been mostly developed on a platform of polycrystalline thin films because of their ease of fabrication. While such thin films have shown remarkable performance in solar cells, the question remains as to what the upper limit of performance might be for perfectly crystalline thin-film devices, and whether this transition is required to further expand the applications of MHP into areas such as optical communications that require semiconductor devices including optical transmitters, modulators, and detectors with high switching speeds.

Two important figures of merit for quantifying the intrinsic electrical properties of a semiconductor are charge-carrier mobility and diffusion length. The electrical mobility $\mu$ is defined as the drift velocity attained by a charge carrier per unit of applied electric field, while the diffusion length $L_{\rm D}$ is the average distance a charge carrier moves between generation and recombination. Despite many measurements of $\mu$ and $L_{\rm D}$ in MHPs, no universal agreement on values have been achieved even for the well-studied \mapbifull\ (\mapbi). The discrepancies may in part be attributed to differences in the purity, stoichiometry and morphology of the samples. For example different fabrication routes of perovskites, such as antisolvent one-step spin-coating methods,\cite{sequential_burschka2013sequential,spinsolvent_jeon2014solvent,spinsolvent_xiao2014fast,grain2015high} air-blading techniques,\cite{airblade_ding2019fully} vapor-assisted deposition,\cite{vapour_yin2016vapor,grain2016high} all-vacuum sequential deposition\cite{vacuum_seq_hsiao2016efficient} and vacuum co-evaporation methods,\cite{vaccum_coep_momblona2016efficient} can lead to different types and concentrations of impurities as well as vastly different grain sizes within perovskite thin films.\cite{grain2015high,grain2016high} However significant discrepancies may also be traced back to different ways of measuring $\mu$ and $L_{\rm D}$.\cite{mobility_laura_review} In this work we show that since MHPs are highly luminescent, the reabsorption of photons emitted from the sample can lead to a significant overestimate of $L_{\rm D}$ and that effect is particularly strong in measurements on single crystals. Thus we reconcile the wide range of values for charge recombination parameters and charge diffusion lengths previously reported from MHP single crystals, as well as discrepancies between single crystals and polycrystalline thin films.    

Conventionally, $\mu$ of a semiconductor is extracted from devices via the Hall effect,\cite{hall_dong2015electron,hall_stoumpos2013semiconducting} space-charge limited current,\cite{hall_dong2015electron,sclc_shi2015low,sclc_saidaminov2015high,rizvi2017improved} time of flight measurements\cite{hall_dong2015electron} or field effect transistor characterization.\cite{fet_chin2015lead,fet_zhang2016carrier} However the strong polarizability of MHPs owing to ion migration\cite{ionmig_unger2014hysteresis,ionmig_meloni2016ionic,ionmig_lan2019physics} and the complexity of making suitable contacts to MHPs can complicate the calculation of mobility from the data acquired via these techniques. As a result, a wide range of electrical mobility have been reported for \mapbi. A literature survey of measured mobility for \mapbi\ is given in Table~S1, where it can be seen that reported mobility values for single crystals of \mapbi\ are particularly inconsistent.  

An alternative approach is to use a non-contact method to determine the intrinsic electrical mobility of a material, via techniques such as microwave conductivity\cite{mwc_hutter2015charge,mwc_reid2016grain,mwc_semonin2016limits,mwc_kim2017300} and terahertz (THz) spectroscopy.\cite{mobility_THz_christian2,mobility_THz_becky,mobility_david} These techniques use time-varying electric fields travelling in free space or a waveguide cavity to perturb and probe the response of charge in a material. This is advantageous as the influence of a metallic contact on the material is removed and the high frequency electric fields of microwave or THz probes avoid the complications associated with ion migration, which occurs on a much longer timescale. THz spectroscopy has the added advantage that conductivity can be observed on a sub-100\,fs timescale allowing charge-carrier dynamics to be followed and recombination parameters extracted.\cite{johnston2016hybrid}

Nonetheless, even within the THz spectroscopy regime, the charge-carrier mobility of \mapbi\ thin films at room temperature has been reported to range from 8 to 35\mob~.\cite{mobility_THz_christian1,mobility_THz_christian2,mobility_THz_becky,mobility_laura_review} Owing to the lack of information about the grain size of those films, it is hard to conduct a systematic comparison between those experimental results and with theoretical models. Therefore, one would expect that $\mu$ measured from perovskite single crystals would give much closer agreement with theory and between different experimental studies, because of the absence of grain boundaries. Surprisingly however, the mobility of \mapbi\ single crystals has been reported to cover an even wider range (0.7 to 600\mob) than for thin films.\cite{sclc_shi2015low,mobility_laura_review,mobility_david}

In this work we study the electrical properties of the prototypical MHP \mapbi\ as both single crystals and polycrystalline thin films and compare direct experimental measurements of $\mu$ with ab initio calculations of transport coefficients based on the Boltzmann transport equation (BTE) and $GW$ quasiparticle band structures. The measured temperature dependence of the single crystal mobility shows close agreement with mobility calculated by solving the BTE.\cite{Ponce2020} A significant difference of the mobility temperature dependence is observed between polycrystalline and single-crystal morphologies, which we attribute to charge-carrier scattering from crystallographic grain boundaries. Indeed, by expanding our BTE theory to include such grain-boundary scattering we are able to replicate all our experimental data. Furthermore, we reconcile the large variation in reported values of $\mu$, $L_{\rm D}$ and charge recombination constants for single crystal MHPs by accounting for the significant influence of photon reabsorption. Thus we have performed a direct experimental comparison of the electrical properties of single-crystal and polycrystalline MHPs, which has allowed us to develop a wholistic model that predicts device-specific parameters such as $\mu$ and $L_{\rm D}$ of MHPs in a range of morphologies over a wide temperature range. Unlike most conventional semiconductor such as Si and GaAs we find that the transition from single crystals to polycrystalline films results in remarkably little degradation to the electrical properties of MHPs, which indicates highly benign grain boundaries in these materials.

\begin{figure}[ht!]
	\includegraphics[width=\textwidth]{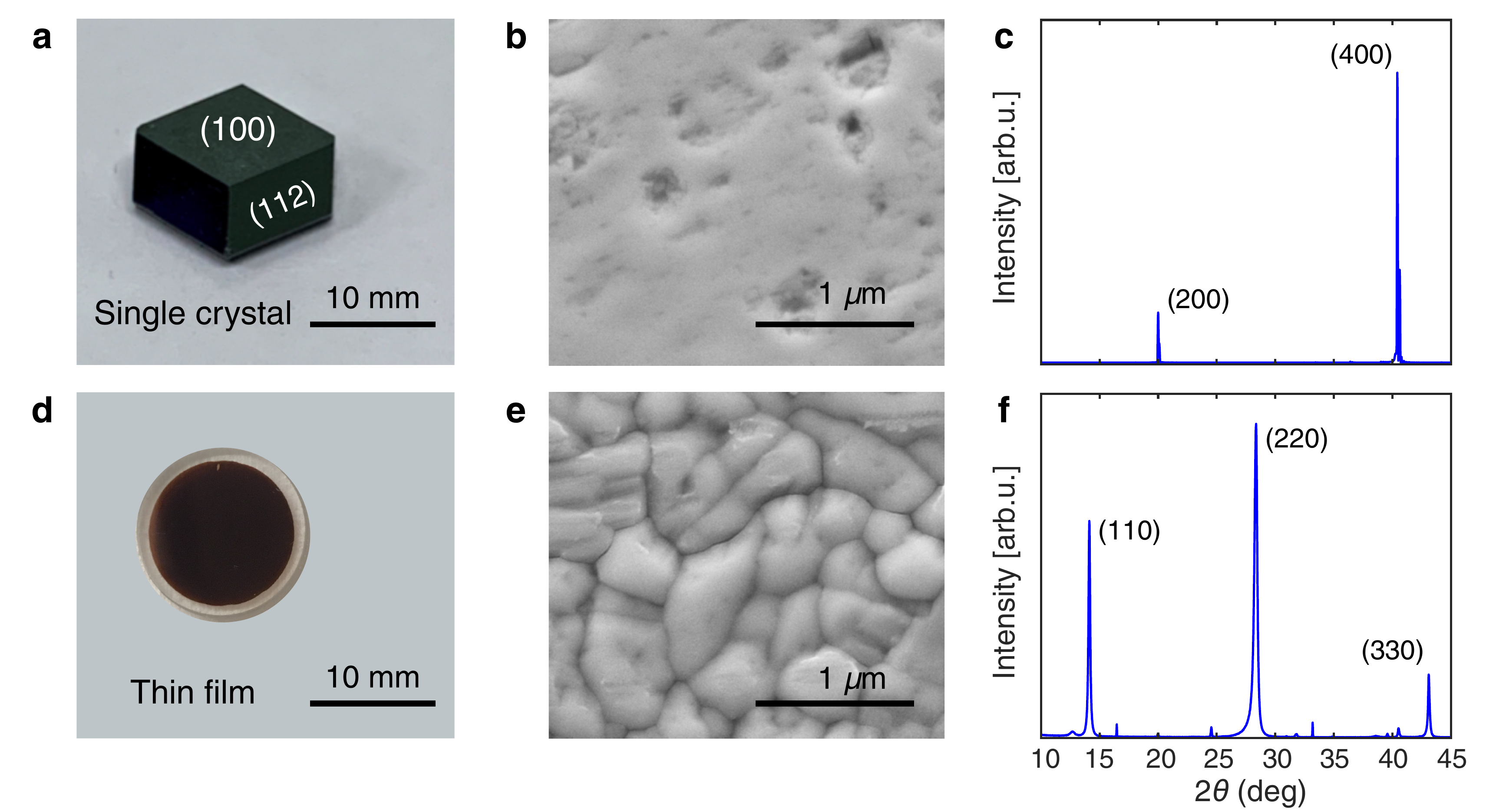}
	\centering
	\caption{(a) Photograph of the \mapbi\ single crystal. (b) SEM image of the single crystal. (c) XRD spectrum of the single crystal's (100) facet. (d) Photograph of the \mapbi\ thin film. (e) Top-down SEM image of the thin film. (f) XRD spectrum of the thin film. The indices shown in panels (c) and (f), e.g. (200) and (110), represent the typical lattice planes observed in \mapbi\ assigned according to previously reported XRD pattern in tetragonal phase.\cite{Baikie2013}}
	\label{figure:xrd_sem}
\end{figure}

%\section{Results and Discussions}
To determine the fundamental upper limit to the mobility of \mapbi\ and hence answer the questions of the influence of grain boundaries and what might be the upper performance limit of perfectly crystalline thin film devices, we chose to study the electronic properties of single crystal samples and compare them directly to those of high-performance polycrystalline thin films. \mapbi\ single crystals were grown using the inverse temperature crystallization technique (see Methods for details of sample growth). A photograph of one of the crystals is shown in Figure~\ref{figure:xrd_sem}a showing large (10\,mm$\times$10\,mm) optically flat facets. A scanning electron micrograph (SEM) of the (100) facet is displayed in Figure~\ref{figure:xrd_sem}b. 

We also grew high-quality \mapbi\ polycrystalline thin films on quartz substrates using vapor co-deposition (see Methods for details of sample growth). It is important to study thin films that are representative of those used in high-efficiency devices, thus we used the same deposition parameters utilized to produce single-junction solar cells with over 19\% PCE.\cite{patel2020light} One of the highly reproducible 600\,nm-thick films used in this study is shown in Figure~\ref{figure:xrd_sem}d and featured an optically flat surface, as is typical of vapor-co-deposited films. The SEM image displayed in Figure~\ref{figure:xrd_sem}e reveals a dense assembly of grains, while a more detailed analysis of the grain size distribution is presented in the Supporting Information (see Figure~S12), which gives a mean length-scale of $\sim580$\,nm. It should be stressed that this value represents an upper limit of grain size as internal misorientation and strain are not revealed by SEM analysis,\cite{EBSDgrain2019local} and smaller crystals may also be present below the observed surface. This suggests that the grain size obtained from the SEM measurement places an upper limit of the real grain size of polycrystalline thin films. Another method to assess crystal size is to examine Scherrer broadening in X-ray diffraction (XRD) peaks resulting from the finite size of the small crystallites. As expected, the XRD peaks measured from the thin film (Figure~\ref{figure:xrd_sem}f) are significantly broadened compared with those of the single crystal (Figure~\ref{figure:xrd_sem}c), with the Scherrer equation returning a crystallite size for the \mapbi\ thin film of $\sim30$\,nm. This value represents a lower limit of crystal size as the contributions of strain and disorder to broadening an XRD peak width are not included in the Scherrer equation.\cite{XRDgrain2016some} Thus the true lateral extent of crystals in the polycrystalline thin films is expected to be $\sim100$\,nm and within the bounds 30--580\,nm. 

\begin{figure}[ht!]
	\includegraphics[width=0.5\textwidth]{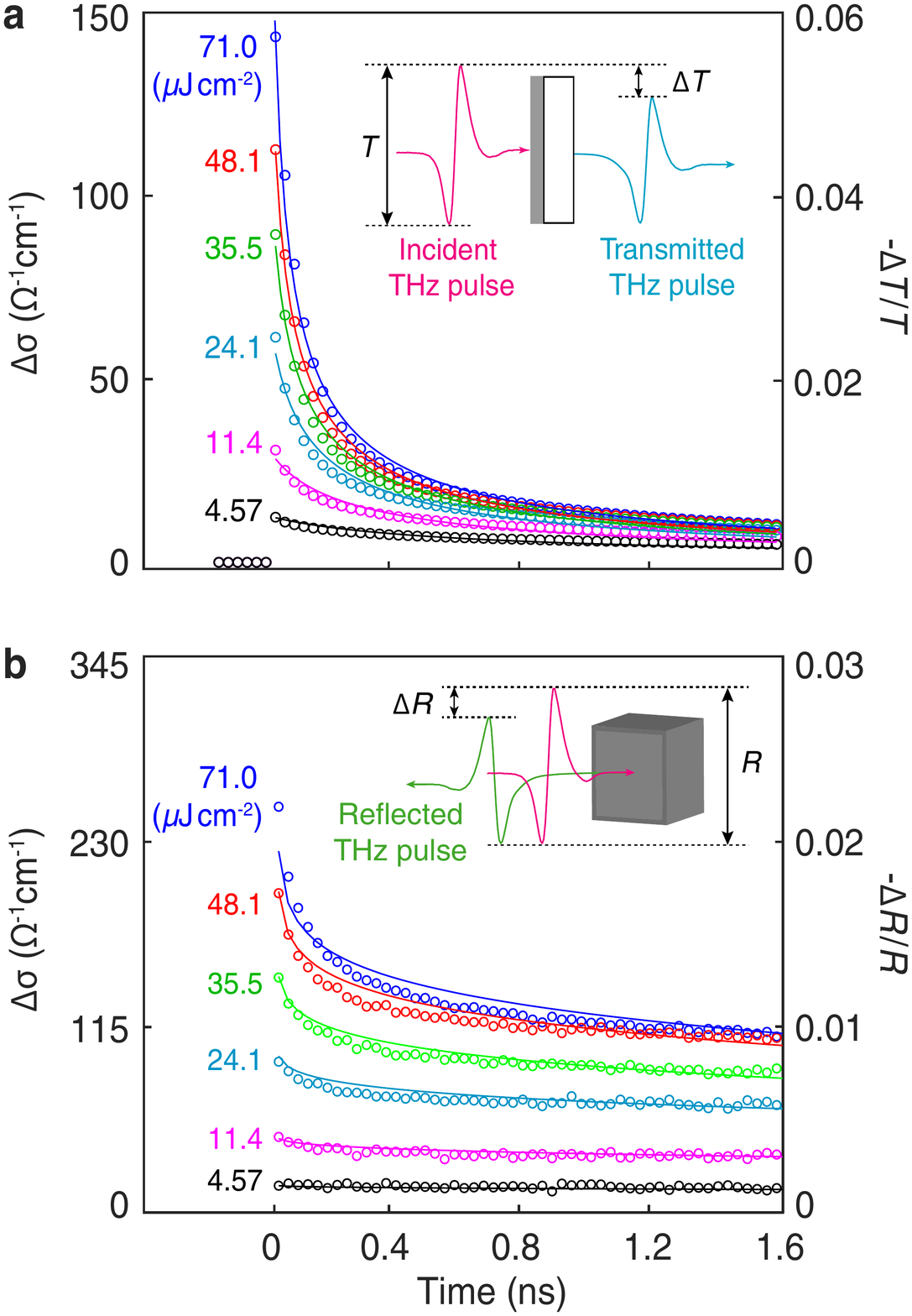}
	\centering
	\caption{Photoconductivity decay dynamics of \mapbi\ measured at room temperature. (a) Measurement of \mapbi\ thin film performed in transmission mode. Reflection measurement of \mapbi\ thin film is shown in Figure~S17 of the Supporting Information. (b) Measurement of \mapbi\ single crystal performed on the (100) facet in reflection mode. The insets illustrate the experimental geometry with the photoexcitation and probing THz pulses incident normally on the sample surface. The circles represent the experimental data and the solid curves represent the theoretical fits. The numerical values of photoconductivity, $\Delta\sigma$, were determined from the measured $-\Delta{T}/T$ and $-\Delta{R}/R$ using Equations~(S4) and (S20) given in the Supporting Information.}
	\label{figure:pumpscan}
\end{figure}

The electrical mobility and charge recombination dynamics of thin-film and single-crystal samples were recorded using the technique of optical-pump-terahertz-probe spectroscopy (OPTPS). The samples were photoexcited by short (35\,fs) pulses of blue light (photon energy 3.1\,eV, central wavelength 400\,nm) and photoconductivity-probed with a sub-picosecond THz pulse. Figures~\ref{figure:pumpscan}a and \ref{figure:pumpscan}b show the room-temperature photoconductivity of thin-film and single-crystal \mapbi\ respectively, as a function of time after photoexcitation by laser pulses with fluences ranging from 4.6 to 71\fluence. The combined electron and hole mobility, $\mu_{\rm e} + \mu_{\rm h}$, at room temperature was found to be $(33\pm2)$\mob\ for the thin film and $(59\pm3)$\mob\ for the single crystal by applying Equations~(S15) and (S23) given in the Supporting Information to the photoconductivity data recorded immediately after photoexcitation (i.e. at $\mathrm{Time}=0$\,ns in Figures~\ref{figure:pumpscan}a and \ref{figure:pumpscan}b). A detailed explanation of how mobility was determined from the raw experimental data is provided in the Supporting Information. Since the mobility was measured at the peak of the photoconductivity decay curve, prior to the charge recombination, diffusion and photon reabsorption, the experimental mobility is independent of those processes (however, when analyzing the charge-carrier decay dynamics after $t=0$, it is crucial to take into account the effect of such charge-carrier diffusion and photon reabsorption as will be discussed later). While the electrical mobility of the thin film is almost half that of the single-crystal \mapbi\ sample, it does not show the three order-of-magnitude drop in mobility seen between single-crystal and polycrystalline GaAs,\cite{GaAs_sc_wolfe1970electron,GaAs_poly_yang1980electrical} indicating that grain boundaries may be more benign in MHPs. Moreover, since the THz measurements are sensitive to the surface conditions, the presence of surface defects on the single crystal can result in an underestimated mobility value. As will be shown later, the experimentally measured mobilities of \mapbi\ single crystal are found to be smaller than the theoretical values calculated by the BTE, which is attributed to the effect of surface defects and impurities in the single crystal.

Understanding charge carrier recombination dynamics is important for modeling semiconductor devices, for example allowing prediction of solar cell PCEs, laser threshold currents and transistor switching times. The three primary mechanisms by which electrons and holes can recombine in MHPs is by (i) Shockley-Read-Hall (SRH) (ii) bimolecular and (iii) Auger recombination.\cite{herz2016charge} SRH recombination is usually a non-radiative process which is mediated by defects or traps. This process is parasitic to the performance of many devices, for example leading to a loss of collected photocurrent in solar cells, and higher threshold currents in lasers. Thus, the aim is often to minimize the SRH recombination path. The SRH process is proportional to the defect density and scales linearly with the charge-carrier density, so it is strongly influenced by the purity of a sample and is the dominant recombination pathway for low charge-carrier densities. The other two processes, bimolecular and Auger recombinations scale quadratically and as the cube of the charge-carrier density, hence are mainly influenced by underlying periodic crystal structure and so depend less on the presence of defects.\cite{johnston2016hybrid} Bimolecular recombination is the recombination of an electron-hole pair and is generally a radiative process in direct bandgap MHPs and should ideally dominate if creating efficient LEDs or lasers. Auger recombination is only significant at very high charge-carrier densities and is parasitic for most optoelectronic devices. As the dominant mechanism by which charges recombine is strongly dependent on charge-carrier density, how charge density changes after the sudden injection of a high density of electron-hole pairs allows the significance of each of these three recombination mechanisms to be quantified. Hence photoinjecting charge and observing charge recombination via the time evolution of photoconductivity is a powerful tool for quantifying key charge-recombination parameters.\cite{johnston2016hybrid}

The evolution of photoconductivity in polycrystalline and single-crystal \mapbi\ as a function of time after photoexcitation (and hence charge density) are displayed in Figures~\ref{figure:pumpscan}a and \ref{figure:pumpscan}b respectively. Observation of such decays in photoconductivity after pulsed photoexcitation allows charge recombination constants for the SRH, bimolecular and Auger mechanisms to be determined. While these recombination constants are important by themselves for device modeling, they also allow the diffusion length $L_{\rm D}$ of charge carriers in the material to be determined for any charge-carrier density in the semiconductor.\cite{johnston2016hybrid} As expected, the decay of photoconductivity for thin-film \mapbi\ is shown in Figure~\ref{figure:pumpscan}a and is seen to depend strongly on the photoexcitation fluence, i.e. the density of photoinjected electron-hole pairs. This phenomenon has been observed previously and rate equations used to extract the recombination constants.\cite{mobility_THz_christian1,mobility_THz_becky,tim_photonreabsorb}

In contrast to the thin-film data, the decay of photoconductivity seen in Figure~\ref{figure:pumpscan}b for the single crystal appears quite different, despite the excitation conditions being identical. Such behavior in single crystals has been observed before, and was attributed to significantly higher values of charge-carrier lifetime, long diffusion lengths\cite{hall_dong2015electron,sclc_shi2015low} and hence potentially much better device performance. However we show that the fundamental recombination parameters underlying these decay curves are remarkably similar between thin film and single crystal despite the large differences between the data in Figures~\ref{figure:pumpscan}a and \ref{figure:pumpscan}b. Indeed, the extended photoconductivity decay curves of single-crystal \mapbi\ shown in Figure~\ref{figure:pumpscan}b arise primarily from photon reabsorption, a process where the photons generated in the radiative bimolecular recombination process are reabsorbed by the sample,\cite{patel2020light} giving rise to new electron-hole pairs, thus extending the observed photoconductivity decay. The effect of photon reabsorption on the charge-carrier dynamics depends strongly on the sample thickness.\cite{tim_photonreabsorb} The thicker the sample is, the more likely the photons are to be reabsorbed before escaping the material, thereby prolonging the decay of photoconductivity. Consequently, photon reabsorption can prolong the decay of photoconductivity in optically thick samples. 

Therefore, to determine accurately the charge recombination constants for thin-film and single-crystal \mapbi\ from experimental data it is necessary to consider how photon reabsorption and charge diffusion affect the data. We used the approach of Crothers et al.\cite{tim_photonreabsorb} to model charge-carrier density, $n(t,z)$, in \mapbi\ as a function of time, $t$, after photoexcitation and depth, $z$, from the sample surface. The measured photoconductivity $\Delta\sigma$ was related to $n(t,z)$ via $\Delta\sigma(t)= \mu e\int_{z=0}^{z_\mathrm{max}} n(t,z)dz$  where $ z_\mathrm{max}$ is the inverse of the absorption coefficient of the sample at the laser wavelength (400\,nm). We modeled both the thin-film and thick single-crystal data in Figure~\ref{figure:pumpscan} with the rate equation,
\begin{equation}
	\frac{\partial{n}}{\partial{t}}=D\frac{\partial{^2n}}{\partial{z^2}}+G-k_1n-k_2n^2-k_3n^3,
	\label{eq:rate}
\end{equation}
where $k_1$ represents the mono-molecular charge recombination rate which is mainly a result of SRH trap-mediated recombination in MHPs; $k_2$ represents the radiative bimolecular recombination constant; and $k_3$ is the Auger recombination constant. $D$ is the diffusion coefficient determined by the charge-carrier mobility and $G$ is the charge-generation rate, which not only includes the charge carriers generated by initial photoexcitation, but also the charges generated by photon reabsorption after radiative bimolecular recombination.\cite{tim_photonreabsorb} The circles in Figure~\ref{figure:pumpscan} represent the experimental data while the solid curves are global fits to Equation~\eqref{eq:rate}. Full details of the model which includes a one-dimensional finite-difference-time-domain solver can be found in the Supporting Information.

The photoconductivity dynamics over the time and fluence range of the experiments presented in Figure~\ref{figure:pumpscan} are expected to be dominated by bimolecular (radiative) recombination, and the large contrast in photoconductivity decay between the single crystal and thin film data at first sight indicates a large difference in their radiative recombination. However, on applying the full rate-equation model given in Equation~\eqref{eq:rate}, it is clear that the slower photoconductivity decay of the thick single crystal arises primarily from photon reabsorption, with the underlying bimolecular recombination being remarkably similar to that of the thin film. The extracted bimolecular recombination constant of the single crystal was $k_{2,\mathrm{crystal}}=8.7\times10^{-10}$\,cm$^3$s$^{-1}$ which is of the same order of magnitude as that of the thin film, $k_{2,\mathrm{film}}=2.6\times10^{-10}$\,cm$^3$s$^{-1}$ (see Table~2 in the Supporting Information for more details). The physical origin of the small drop in the bimolecular recombination constant in the thin film compared with the single crystal is likely to be associated with a small degree of electron-hole separation at grain boundaries in the thin films. Thus we find that the bimolecular recombination in single-crystal \mapbi\ is consistent with the thin-film value measured in this study and also with previously reported values of other \mapbi\ thin films.\cite{mobility_THz_becky,tim_photonreabsorb}

One of the most dramatic differences between thin-film and single-crystal \mapbi\ has been the much larger charge-carrier diffusion length $L_{\rm D}$ observed for the thick single crystals, which is reported to differ by a few orders of magnitude.\cite{hall_dong2015electron} However after accounting for the effects on photon reabsorption in our measurements we find a much smaller difference between $L_{\rm D}$ values for single-crystal and polycrystalline \mapbi. We define the diffusion length as the average distance traveled by a charge carrier between generation and recombination/trapping at a uniform carrier density $n$ in the absence of an electric field,
\begin{equation}
	L_{\rm D}(n)=\sqrt{\frac{D}{R(n)}},
\end{equation}
where $R(n)=n^2k_3+nk_2+k_1$ is the total recombination rate and $D={\mu}k_{\rm B} T/e$ is the diffusion constant. This function can thus be determined using experimentally determined values of mobility $\mu$ and recombination constants $k_1$, $k_2$ and $k_3$. Using the mobility values measured at $t=0$, $k_2$ values extracted from the photon reabsorption model described by Equation~\eqref{eq:rate}, and $k_1$ values obtained from time-resolved photoluminescence (TRPL) measurements (see Figure~S13 in the Supporting Information), the diffusion lengths for the thin film and the single crystal are 1\micron\ and 2.83\micron\ respectively for a charge-carrier density consistent with 1 sun illumination (1kWm$^{-2}$ AM1.5-filtered light). $L_{\rm D}$ is plotted as a function of charge-carrier density for thin-film and single-crystal \mapbi\ in Figure~S15 in the Supporting Information. As discussed in the Supporting Information, if photon reabsorption is neglected in the determination of $k_2$ then $L_{\rm D}$ is artificially longer, as it includes on average more than one generation-recombination events and will be dependent on the thickness of the sample. Such overestimate of the single crystal's diffusion length is particularly significant at low values of $k_1$ ($<10^7$\,s$^{-1}$) where the bimolecular recombination becomes more prominent. Therefore, an inaccurate determination of the bimolecular recombination constant will lead to an unrealistically long diffusion length for the single crystal. These results indicate the importance of stating charge-carrier density as well as taking into account photon reabsorption when reporting diffusion lengths for \mapbi\ and direct bandgap semiconductors generally. Meanwhile, it is worth noting that some other factors may also give rise to different diffusion lengths observed in single crystals and thin films, such as capacitive charging and discharging effects, defects and impurities present in the samples.

In order to gain a better understanding of fundamental limits to intrinsic electrical mobility and charge-carrier diffusion length in \mapbi\ we solved for the orthorhombic phase of \mapbi\ the ab initio BTE\cite{Ponce2020} in the self-energy relaxation time approximation\cite{Ponce2018} including spin-orbit coupling effects and with electronic states computed with an eigenvalue self-consistent many-body $GW$ method,\cite{Sangalli2019} vibrational eigenstates from density-functional perturbation theory (DFPT)\cite{Gonze1997a,Baroni2001,Giannozzi2017} where the scattering rates were obtained through a maximally localised Wannier function interpolation\cite{Pizzi2020} of the first-principles electron-phonon matrix elements.\cite{Giustino2007,Ponce2016a,Giustino2017} Details of the calculations are given in the Methods and in a prior work where we reported the averaged mobility $\mu=(\mu_{\rm e}+\mu_{\rm h})/2$.\cite{mobility_DFT_ponce} However, in our THz photoconductivity measurement, the observed mobility is in fact the sum of electron and hole mobilities. Therefore, using the room-temperature electron and hole mobilities calculated for perfect, detect-free single-crystal \mapbi\ where $\mu_{\rm e}=33$\mob\ and $\mu_{\rm h}=50$\mob, a total theoretical electrical mobility $\mu=\mu_{\rm e}+\mu_{\rm h}=83$\mob\ is obtained. This value should represent an upper limit of any measurement on an imperfect real crystal, which will contain impurities, defects and surfaces. Thus the value of mobility measured on our real crystal $(59\pm3)$\mob\ agrees well with the value calculated using the first-principles technique. We note that our computed mobility is in line with the one computed in Ref.\citen{mayers2018} but lower than Refs.\citen{wu2014temperature,wright2016electron,mobility_temp_Frost,herz2018lattice,lacroix2020} which neglected the role of multi-phonon Fr\"ohlich coupling as discussed in Ref.\citen{Schlipf2018}. This is crucial since it was shown that at least two sets of modes contribute to the mobility of halide perovskites.\cite{mobility_DFT_ponce}

Specifically, we find that charge-carrier scattering (and hence mobility) in \mapbi\ is primarily influenced by three longitudinal optical (LO) phonon modes: (i) a Pb--I--Pb bending with a 4.3\,meV energy, (ii) a dominant Pb--I stretching mode at 14.4\,meV that accounts for half of the scattering and (iii) a broad contribution around 21\,meV originating from the librational modes of the CH$_3$NH$_3$ molecules. Thus including multiple phonon modes in calculations of charge-carrier scattering reconciles the overestimate of theoretical mobility in \mapbi\ compared with experimental values.  

\begin{figure}[ht!]
	\includegraphics[width=\textwidth]{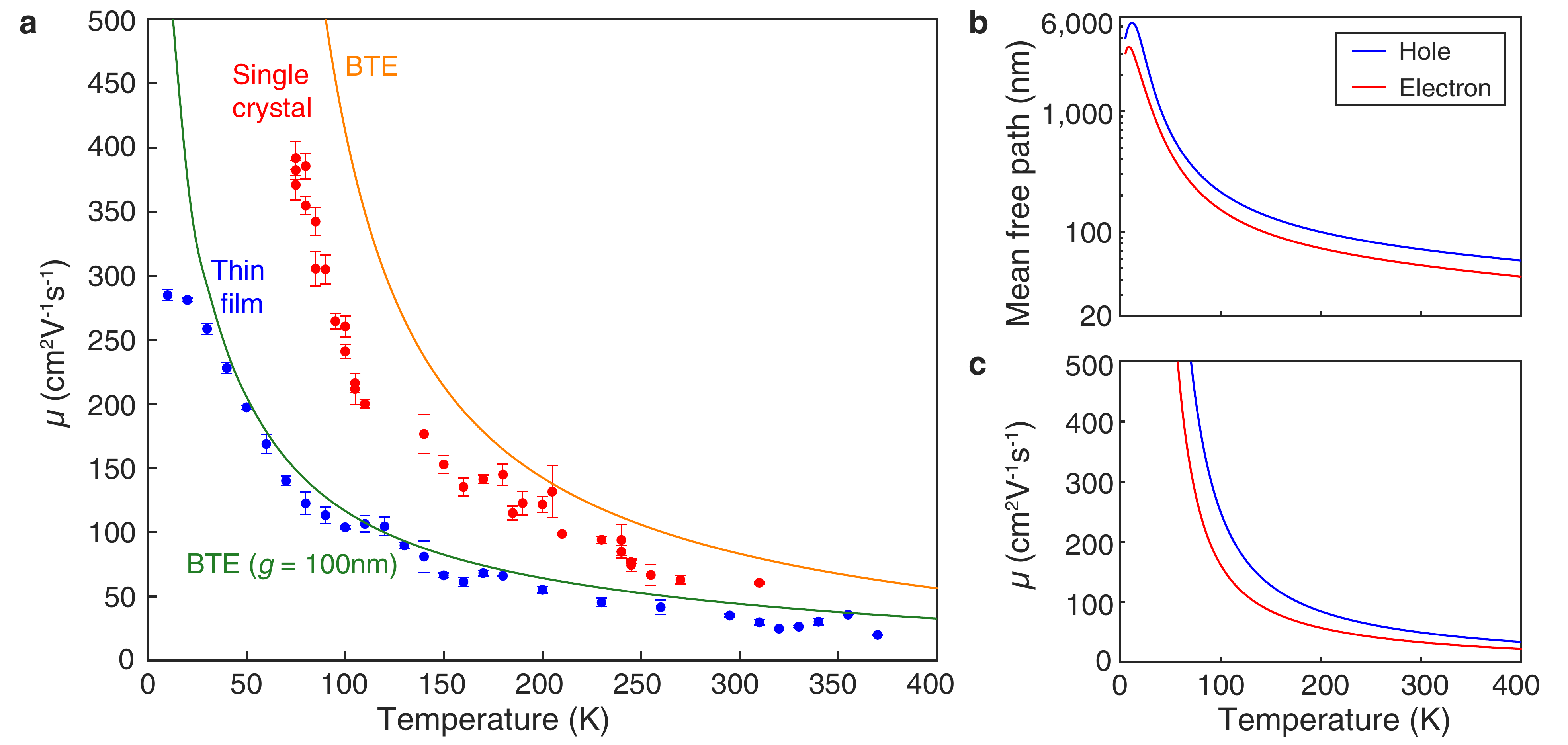}
	\centering
	\caption{(a) Charge-carrier mobility of \mapbi\ as a function of temperature, where the blue and red circles represent the experimental data of thin film and single crystal respectively. Each data point was measured repeatedly three times, from which the error bar was determined by the standard deviation. The orange line represents the ab initio BTE calculation of intrinsic phonon-limited mobility where all the phonon modes are included, whilst the green line represents the BTE calculation including grain-boundary scattering with crystal size $g=100$\,nm. (b) Mean free path of pure \mapbi\ single crystal in orthorhombic phase. (c) Intrinsic mobility of \mapbi\ single crystal obtained by solving the BTE in the self-energy relaxation time approximation.}
	\label{figure3:mu_mfp}
\end{figure}

To test the validity of our multiple-phonon-mode BTE model we compared measured and calculated values for electrical mobility of \mapbi\ over a wide temperature range. Figure~\ref{figure3:mu_mfp}a displays the measured electrical mobility of single-crystal \mapbi\ (red circles) over a temperature range 75--310\,K in comparison to our multi-phonon-mode BTE calculations (orange line). As temperature is lowered, mobility increases which is consistent with reduced occupancy of phonon modes, and hence reduced electron-phonon Fr\"ohlich coupling. 

Experimental mobility data are often fitted with a temperature power law relation to help determine the physical origin of scattering.\cite{mobility_laura_review,herz2018lattice} However, simple power law models of LO-phonon coupling assume only one LO phonon branch and simplified dispersion relations, which are clearly not applicable for \mapbi\ which has many atoms in its primitive unit cell basis, and hence a dense phonon spectrum. Thus the only way to properly account for the temperature dependence of electrical mobility in MHPs is to consider the occupancy and Fr\"ohlich coupling of all significant phonon modes as a function of temperature, as we have done with our BTE model.

The temperature dependence of electron and hole mobilities calculated by solving the BTE for a pure \mapbi\ crystal without any impurities is shown in Figure~\ref{figure3:mu_mfp}c, while the combined electron and hole mobility (orange line) is compared with the experimental mobility data (red circles) in Figure~\ref{figure3:mu_mfp}a. We expect that the calculated mobility will represent an upper limit for the mobility of any real crystal, owing to the presence of impurities in real crystals. Indeed the experimental single crystal data (red circles) agrees well with the shape of the theoretical data and is bounded by it. The overestimate is somewhat higher at lower temperatures, where ionized impurity scattering, which is not included in the BTE calculations, becomes more significant.  

In contrast, the values of experimental mobility are significantly lower for the polycrystalline thin film (blue circles in Figure~\ref{figure3:mu_mfp}a) compared with the single crystal and importantly the discrepancy becomes larger at lower temperature. We hypothesize that this difference is related to presence of grain boundaries in the polycrystalline thin films, and test this hypothesis by modeling grain-boundary scattering within our BTE framework.  

To gain insight into the influence of grain boundaries on the electrical mobility of polycrystalline \mapbi\ we implemented the model of Mayadas and Shatzkes\cite{Mayadas1970} where the mean free path of charge carriers is limited by the extent of grain boundaries. The model is an extension of the Boltzmann transport theory to include reflection of the charge carriers at the grain boundaries of polycrystalline thin films and can be written as\cite{Mayadas1970}
\begin{equation}\label{eq:grain}
	\mu_{\rm film}=\mu_{\rm phonon}\left[1-\frac{3}{2}\alpha+3\alpha^2-3\alpha^3\ln(1+\frac{1}{\alpha})\right],
\end{equation} 
where $\alpha=R(\lambda/g)/(1-R)$, $\lambda$ is the carrier mean free path, $R$ is the probability of reflection at the grain boundary and $g$ is the grain size. $\mu_\mathrm{film}$ and $\mu_\mathrm{phonon}$ represent the electrical mobility calculated by the BTE model with and without the consideration of grain-boundary scattering respectively. Following previous works, we set the probability of reflection at $R=0.5$\cite{Steinhoegl2002,Sun2010} and the carrier mean free path is computed ab initio by direction-averaging the carrier mean free path weighted by a transport occupation,
\begin{equation}
	\lambda_{\rm transport} = \frac{\sum_{\alpha,n\boldsymbol{k}} w_{\boldsymbol{k}} f_{n\boldsymbol{k}} (1-f_{n\boldsymbol{k}}) |\textbf{v}_{n\boldsymbol{k}}^\alpha| \tau_{n\boldsymbol{k}}}{3 \sum_{n\boldsymbol{k}} w_{\boldsymbol{k}} f_{n\boldsymbol{k}} (1-f_{n\boldsymbol{k}}) },
\end{equation}
where $w_{\boldsymbol{k}}$ is the $\boldsymbol{k}$-point weight, $f_{n\boldsymbol{k}}$ is the Fermi-Dirac occupation function, $\mathbf{v}_{n\boldsymbol{k}}$ is the carrier velocity and $\tau_{n\boldsymbol{k}}$ is the electron-phonon lifetime for a state of band $n$ and momentum $\boldsymbol{k}$. The mean free path average for electrons and holes is shown in Figure~\ref{figure3:mu_mfp}b. For comparison, the mean free path calculated for the most relevant energy $(3/2)k_\mathrm{B}T$ is given in the Supporting Information (see Figure~S9).
The green line in Figure~\ref{figure3:mu_mfp}a shows the ab initio calculated $\mu_{\rm film}$, with crystal grain size $g$ as the only free parameter. A grain size $g= 100$\,nm, which as discussed earlier is reasonable for our samples, was found to be in excellent agreement with experimental mobility data for temperatures above 30\,K. Below 30\,K exciton formation, which is not included in the theoretical model, is expected to dominate and is consistent with the reduced experimental mobility at low temperatures.\cite{Davies2018} The effect of increasing and reducing grain size on mobility is displayed in Figure~S10 in the Supporting Information with mobility being most sensitive to grain size at temperatures below 200\,K. Thus, our BTE theory, when corrected for grain-boundary scattering provides an excellent prediction of electrical mobility in \mapbi\ for polycrystalline and single-crystal morphologies over a wide temperature range. We emphasize that this settles a long-standing debate in the literature whereby acoustic scattering,\cite{karakus2015,yi2016} single optical mode scattering,\cite{filippetti2016} ionized impurity scattering,\cite{zhao2016} piezoelectric scattering\cite{lu2017} or polaronic scattering\cite{mobility_temp_Frost} were all mentioned as contributing to the carrier mobility in \mapbi. Here we have shown that only multi-mode optical-phonon scattering for single crystal augmented by grain-boundary scattering for polycrystalline thin film are sufficient.

In conclusion, we performed a comprehensive experimental and theoretical study of the electrical properties of both single-crystal and polycrystalline \mapbi. We reconcile the large discrepancy in previously published values of key figures of merit such as mobility, diffusion length and recombination parameters by including the effects of photon reabsorption. In particular, we find that neglecting photon reabsorption when modeling thick single crystals leads to a significant overestimate of charge-carrier diffusion length. Our experimental data agree extremely well with ab initio Boltzmann transport calculations when the Fr\"ohlich interaction of multiple phonon modes are included. This result explains the overestimation of mobility in previous calculations where only one phonon branch was included. The BTE model provided excellent agreement with single-crystal mobility data without any fitting parameters. However, the measured mobility of polycrystalline thin films deviated from the single-crystal data and BTE model. We found that grain-boundary scattering accounted for this deviation and included this effect in our BTE model. While the mobility of polycrystalline films is primarily limited by grain-boundary scattering at cryogenic temperatures, we found that room-temperature mobility is dominated by LO-phonon scattering. As such we find the mobility only dropping from 59\mob\ in the single crystal to 33\mob\ for polycrystalline \mapbi\ indicating the benign nature of these grain boundaries at room temperature. Overall this study provides a complete picture of the fundamental electrical properties of the model MHP \mapbi\ in both single-crystal and polycrystalline morphologies. We thus unify apparently contradictory previous experimental and theoretical studies. Our results indicate the polycrystalline thin films possess similar performance to single crystals at the usual operating temperature of electronic device. This is promising for future applications of MHP thin films in high-speed devices such as transistors, emitters, modulators and detectors, as well as for upscaling future generations of  solar cells and lighting panels.

\section{Experimental and Computational Details}
%\threesubsection{Sample preparations}\\
\textit{\mapbi\ Single Crystal Fabrication.} The \mapbi\ perovskite single crystals were prepared by inverse temperature crystallization.\cite{crystalprep2,sclc_saidaminov2015high} \mapbi\ precursor (1.25\,mol\,L$^{-1}$) was prepared by adding PbI$_2$ (461\,mg) and methylammonium iodide which is defined as CH$_3$NH$_3$I (159\,mg) into $\gamma$-Butyrolactone (0.8\,mL), heated at 60\,$^\circ$C for 2 hours with stirring. The precursors were filtered with syringe filters (0.22\,$\mu$m pore size). The obtained solution was transferred to clean containers, which were kept on a stable hot plate and gradually heated to 120\,$^\circ$C and kept for another 6 hours. Crystals were formed at the bottom of the containers. Finally, the crystals were collected and dried at 60\,$^\circ$C in vacuum oven for 12 hours.

\textit{\mapbi\ Thin Film Fabrication.} The \mapbi\ thin films were prepared in two steps. (1) Cleaning of substrates: z-cut quartz substrates were cleaned with hellmanex solution, followed by a thorough rinse with deionised water. The substrates were then washed with acetone, isopropanol, and ethanol. Thereafter the substrates were plasma etched in O$_2$ for 10 minutes. (2) Thermal co-evaporation of \mapbi: the \mapbi\ was fabricated using thermal evaporation as reported previously.\cite{patel2020light} In brief, MAI and PbI$_2$ were placed in separate crucibles, and the substrates were mounted on a rotating substrate holder to ensure that a uniform film was deposited. The temperature of the substrates was kept at 21\,$^\circ$C throughout the deposition. The chamber was evacuated to reach a high vacuum of $10^{-6}$\,mbar, before heating the PbI$_2$ and the MAI. The substrates were then exposed to the vapour. The rates of both the MAI and PbI$_2$ were monitored using a quartz crystal microbalance. The thickness of the perovskite thin film was set by controlling the exposure time of the substrates to the vapor. 

\textit{Optical-Pump-THz-Probe Spectroscopy (OPTPS).} The OPTPS was utilized to measure the photoconductivity and the electrical mobility of \mapbi\ thin films and single crystals. The THz pulse was generated by a THz spintronic emitter due to the inverse spin Hall effect.\cite{spintronic} An amplified ultrafast (35\,fs) laser beam with an average power of 4\,W and central wavelength of 800\,nm was split into three arms: a probe (THz) beam, a gate beam and a pump beam. The THz pulse was detected by a 0.1mm-thick (110) ZnTe crystal together with a Wollaston prism and a pair of balanced photodiodes via electro-optic sampling. The pump beam was converted from 800\,nm to 400\,nm by a $\beta$-barium-borate (BBO) crystal to photoexcite the \mapbi\ thin films and single crystals. Under photoexcitation, the photoinjected charge carriers give rise to a reduction of the THz intensity, which is known as photoconductivity and used for extracting the electrical mobility. Whilst the photoconductivity of \mapbi\ thin film was measured in transmission mode, the photoconductivity of the single crystal was measured in reflection mode instead since little THz signal could transmit through the thick crystal. A schematic of the OPTPS setup in transmission and reflection modes is shown in Figure~S1 in the Supporting Information. All measurements were repeated three times at each temperature, from which the uncertainty was determined by the standard deviation. A detailed derivation of the electrical mobility is given in the Supporting Information.

\textit{Thermometry and Temperature-Dependent PL Spectroscopy.} The very low thermal conductivity of \mapbi\ has been reported to be below 0.5\,W(mK)$^{-1}$,\cite{thermal_pisoni2014ultra,thermal_heiderhoff2017thermal,thermal_gold2018acoustic} which is hundreds of times smaller than that of conventional semiconductors such as GaAs.\cite{thermo_nolas2004thermal} Therefore, extreme care must be taken when making temperature dependent measurements, particularly if samples are exposed to localized heating, such as via laser excitation. For thin-film \mapbi\ deposited on a quartz substrate, the low thermal conductivity is negated by the material's close contact with the quartz substrate, which has a high thermal conductivity over a wide temperature range. Unfortunately for the cm-sized \mapbi\ single crystal in our study the low thermal conductivity makes the experiments extremely challenging, and great care was taken to ensure that the recorded temperature was actually its lattice temperature at the position where the measurements were made. Therefore, we developed a PL-corrected THz technique to measure the mobility of \mapbi\ single crystal with accurate temperature determination. To improve the thermal contact between the single crystal and the cold-finger cryostat (Oxford Instruments, MicrostatHe), which was used to change the crystal temperature in our OPTPS measurement, we inserted a sapphire substrate at the front of the single crystal, so that the heat generated by the pump beam could be dissipated more efficiently, which enabled us to cool the crystal down to 75\,K and observe a clear phase transition in its PL spectrum at 160\,K. In the meantime, to determine the crystal temperature more accurately, we measured the corresponding PL spectrum at different temperatures using both cold-finger and gas-exchange cryostats. In the gas-exchange-cryostat setup (Oxford Instruments, OptistatCF2) which is separate from the OPTPS setup, since the crystal was immersed in helium gas, there was no thermal contact issue any more and the temperature registered by the sensor in the gas-exchange cryostat was the true temperature of the crystal. Therefore, we were able to use the PL spectrum measured by the gas-exchange cryostat as a reference to correct the temperature measured by the cold-finger cryostat. A detailed PL-facilitated temperature-correction process is given in the Supporting Information. In the cold-finger-cryostat setup, the PL spectrum was generated from excitation by the pump beam (400\,nm, 35\,fs) used in the OPTPS setup and collected by a fibre-coupled spectrometer (Horiba Scientific, iHR320) and detected by a CCD (Horiba Scientific, Si Symphony II). In the gas-exchange-cryostat setup, the PL spectrum was generated from excitation by a picosecond pulsed diode laser (PicoHarp, LDH-D-C-405M) at central wavelength of 398\,nm with the signal subsequently collected and coupled into a different spectrometer (Princeton Instruments, SP-2558), and detected by an iCCD (Princeton Instruments, PI-MAX4).

\textit{Computational Methods Based on Density-Functional Theory (DFT).} We performed DFT calculations using pseudopotentials and planewaves, as implemented in the Quantum ESPRESSO package.\cite{Giannozzi2017} We used the local density approximation (LDA) with norm-conserving pseudopotentials from the PseudoDojo repository.\cite{Setten2018} We used fully relativistic pseudopotentials which includes the effect of spin-orbit coupling as well as semicore electrons in the case of Pb. We used a plane-wave kinetic energy cutoff of 100\,Ry and the following orthorhombic lattice parameters $a=8.836$\,\AA, $b=12.581$\,\AA, and $c=8.555$\,\AA.\cite{Baikie2013} We calculated phonons using DFPT\cite{Gonze1997a,Baroni2001} with a 4$\times$4$\times$4 $k$-points and 2$\times$2$\times$2 $q$-points grids. We corrected the DFT band structures via the quasiparticle $GW$ method, using the Yambo code.\cite{Sangalli2019} We employed a higher plane-wave kinetic energy cutoff of 150\,Ry, we evaluated the exchange self-energy and the polarisability using cutoffs of 80\,Ry and 6\,Ry, respectively, and performed the summations over empty states using 1,000\,bands for the calculation of the polarisation and the Green's function. The frequency-dependence of the screened Coulomb interaction was described via the Godby-Needs plasmon-pole model,\cite{Godby1989} using a plasmon-pole energy of 18.8\,eV. Since the DFT gap of lead halide perovskites is very small due to spin-orbit coupling,\cite{Davies2018} we went beyond the $G_0W_0$ approximation by including self-consistency on the eigenvalues. We applied self-consistency by using the strategy of Ref.~\citen{mobility_DFT_ponce} which includes a wavevector-dependent scissor so as to obtain accurate effective masses. The Brillouin zone was sampled via a 4$\times$4$\times$4 unshifted grid, and the termination scheme of Ref.~\citen{Bruneval2008} was employed to accelerate the convergence with respect to the number of empty states. We calculated the electron-phonon matrix elements and scattering rates using the EPW code,\cite{Ponce2016a} in conjunction with the wannier90 library.\cite{Pizzi2020} We included spin-orbit coupling in all calculations. We started from the 2$\times$2$\times$2 grid of phonon wavevectors and interpolated on a fine grid containing 100,000 points and following a $\Gamma$-centred Cauchy distribution weighted by their Voronoi volume. We neglect anharmonic effects which could be important at room temperature and above.~\cite{mayers2018}

\begin{acknowledgement}
This work was funded by the Engineering and Physical Sciences Research Council (EPSRC). M.B.J. thanks the Alexander von Humboldt Foundation for support. S.P. acknowledges the support from the European Union's Horizon 2020 Research and Innovation Programme, under the Marie Sk\l{o}dowska-Curie Grant Agreement SELPH2D No. 839217. F.G.'s contribution was supported as part of the Computational Materials Sciences Program funded by the U.S. Department of Energy, Office of Science, Basic Energy Sciences, under Award DE-SC0020129.
\end{acknowledgement}

%
%\begin{suppinfo}
%	The following files are available free of charge.
%	\begin{itemize}
%		\item Supporting Information
%	\end{itemize}
%\end{suppinfo}

\begin{suppinfo}
Details of THz spectroscopy setup, derivations of charge-carrier mobility from OPTPS measurements, analysis of grain-boundary scattering effect on electrical mobility, analysis of photon reabsorption, and details of PL-facilitated temperature correction technique. 
\end{suppinfo}

\newpage
\providecommand{\latin}[1]{#1}
\makeatletter
\providecommand{\doi}
  {\begingroup\let\do\@makeother\dospecials
  \catcode`\{=1 \catcode`\}=2 \doi@aux}
\providecommand{\doi@aux}[1]{\endgroup\texttt{#1}}
\makeatother
\providecommand*\mcitethebibliography{\thebibliography}
\csname @ifundefined\endcsname{endmcitethebibliography}
  {\let\endmcitethebibliography\endthebibliography}{}

\end{document}